\documentclass[elsarticle]{elsarticle}

\usepackage{lineno,hyperref}
\modulolinenumbers[5]

\journal{Physics Letters B}

\begin{document}

\begin{frontmatter}

\title{Off-shell initial state effects, gauge invariance and angular distributions in the Drell-Yan process}


\author[mymainaddress,mysecondaryaddress]{Maxim Nefedov}

\author[mysecondaryaddress]{Vladimir Saleev\corref{mycorrespondingauthor}}
\cortext[mycorrespondingauthor]{Corresponding author}

\address[mymainaddress]{Samara National Research University, Moskovskoe Shosse 34, 443086 Samara, Russia}
\address[mysecondaryaddress]{II. Institut f$\ddot{u}$r Theoretische Physik, Universit$\ddot{a}$t Hamburg,
Luruper Chaussee 149, 22761 Hamburg, Germany}

\begin{abstract}
We discuss production of Drell-Yan lepton pairs at hadron colliders
in the framework of the parton Reggeization approach, which includes
off-shell initial state effects in a gauge-invariant way. Other
possible prescriptions to restore gauge-invariance of
hard-scattering coefficient with off-shell initial-state partons are
also investigated and significant differences for the resulting
structure functions are found, especially for the
$F_{UU}^{(\cos2\phi)}$. We compare our numerical results for
$q_T$-spectra of the lepton pair with experimental data, obtained by
E-288 collaboration ($\sqrt{S}=19.4$ and 23.8 GeV) and find a good
agreement. Also we perform predictions for the Drell-Yan structure
functions at NICA $pp$-collider ($\sqrt{S}=24$ GeV).
\end{abstract}


\end{frontmatter}


\section{Introduction}
\label{intro} The Drell-Yan(DY) process of production of lepton
pairs with large invariant mass in hadronic collisions is one of the
most important tests of perturbative quantum chromodynamics (QCD),
as well as the unique source of information about partonic structure
of hadrons. Apart from the inclusive cross-section, differential
w.r.t. squared invariant mass ($Q^2$), transverse momentum ($q_T$)
and rapidity ($Y$) of the lepton pair or some equivalent variable,
such as momentum fraction in the Collinear Parton model (CPM)
($x_{A,B}$), also structure functions or angular coefficients, wich
parametrize the angular distribution of leptons in the rest frame of
the lepton pair are often under consideration. Behavior of the
latter class of observables in the region of relatively small
$q_T\leq Q$ will be the main subject of the present paper.

  At low $q_T\ll Q$, already the prediction of inclusive cross-section, integrated over all directions of
lepton momentum in the center-of-mass frame of the pair, presents a
considerable difficulty for the conventional CPM, since at any fixed
order of perturbation theory the cross-section diverges as $1/q_T^2$
at $q_T\to 0$. These un-physical divergence is regulated through the
resummation of higher-order corrections in $\alpha_s$ enhanced by
$\log^2 (q_T/Q)$ and $\log(q_T/Q)$ through Collins-Soper-Sterman
formalism~\cite{CSS}, which later has been reformulated in a form of
Transverse Momentum Dependent (TMD) factorization
theorem~\cite{Collins}.

  In TMD-factorization, the hard-scattering coefficient (HSC) doesn't depend explicitly on the transverse momenta of
colliding partons. Instead, it is calculated with on-shell
initial-state partons and corresponding partonic tensor
automatically satisfies the QED Ward identity. However, it is
possible to develop a complementary approach to TMD factorization,
starting not from the collinear limit but from Multi-Regge limit for
QCD scattering amplitudes, i.e. from the limit when all final-state
particles are highly-separated in rapidity. We call such scheme of
calculations -- the Parton Reggeization Approach (PRA). It's logic
is outlined below in the Sec.~\ref{PRA}.

  In PRA, the HSC, although being gauge-invariant, nevertheless explicitly depends on the transverse momenta of initial-state partons.
Below we demonstrate, that this dependence is important for the
calculation of the angular structure functions, since alternative
prescriptions which one could propose to naively ``restore'' the
gauge-invariance of HSC with off-shell initial-state partons lead to
significantly different numerical values for them. Most importantly,
for the structure function $F_{UU}^{(\cos 2\phi)}$ the difference
starts already at leading power in $q_T/Q$.

  The present paper has following structure: In Sec.~\ref{PRA} main ideas of Parton Reggeiztion Approach are outlined.
In Sec.~\ref{spectra} the analytic results for angular structure
functions in PRA  are listed. The same quantities in the alternative
gauge-invariant TMD-factorization scheme, which we call
quasi-on-shell scheme are derived in the Sec.~\ref{QOS}, and
in the Sec.~\ref{num} the numerical results for structure functions
in both schemes are presented. We perform our numerical computations
for the planned energy of $pp$-collisions at NICA collider:
$\sqrt{S}=24$ GeV. Comparison with experimental data of E-288
Collaboration for a very close energies, is also presented in the
Sec.~\ref{num} to justify the extension of PRA to this domain of
relatively low energies.

\section{Parton Reggeization Approach}
\label{PRA} More detailed introduction to the PRA and derivation of
our factorization formula is presented in the Ref.~\cite{PRA:BBbar}.
Here we only briefly outline the main ideas. Factorization formula
of PRA is based on {\it modified-MRK approximation} for QCD matrix
elements. This approximation smoothly interpolates between
well-known collinear and Multi-Regge asymptotics (see e.g.
Ref.~\cite{MRK-rev} for the review of the latter) of matrix element
of ordinary CPM hard subprocess with emission of two additional
partons. In the collinear limit, additional partons have $|{\bf
k}_T|\ll \mu$ where $\mu$ is the hard scale ($\mu\sim Q$ in the case
of DY process), while in the Multi-Regge limit, additinal partons
are highly separated in rapidity from the system of interest
($l^+l^-$ for DY process), while their typical $|{\bf k}_T|\sim
\mu$. In both limits, QCD matrix element can have the
$t$-channel-factorized form, however in the MRK case the partons,
propagating in the $t$-channels are not ordinary QCD quarks and
gluons, but special gauge-invariant degrees of freedom of
high-energy QCD, called Reggeized quarks ($Q$) and gluons ($R$). Due
to the $t$-channel-factorized form of the mMRK-approximation, the
cross section of lepton pair production in proton-proton collisions,
$p(P_1)+p(P_2)\to l^+(k_1)+l^-(k_2)+X$, can be presented in
$k_T-$factorized form:
 \begin{equation}
  d\sigma = \int\limits_0^{1}\frac{dx_1}{x_1} \int \frac{d^2{\bf q}_{T1}}{\pi} {\Phi}_q(x_1,t_1,\mu^2)
  \int\limits_0^1 \frac{dx_2}{x_2} \int \frac{d^2{\bf q}_{T2}}{\pi} {\Phi}_{\bar{q}}(x_2,t_2,\mu^2)\cdot d\hat{\sigma}_{\rm
  PRA},\label{eq_factor}
\end{equation}
where $x_{1}=q_1^+/P_1^+$, $x_2=q_2^-/P_2^-$, four-momenta of
partons in the initial-state of the leading order (LO) PRA
hard-scattering subprocess $Q(q_1)+\bar{Q}(q_2)\to l^+ + l^-$ are
parametrized as $q_1=\frac{1}{2} q_1^+ n_- + q_{T1}$,
$q_2=\frac{1}{2} q_2^- n_+ + q_{T2}$, $t_{1,2}={\bf
q}_{T1,2}=-q_{1,2}^2$, and light-cone vectors are defined as
$n_-^\mu=2 P_1^\mu/\sqrt{S}$, $n_+^\mu=2 P_2^\mu/\sqrt{S}$ where
$S=(P_1+P_2)^2=2P_1P_2$. For any four-vector the light-cone
components are $k^{\pm}=\left(kn^{\pm}\right)$, so that
$k^2=k^+k^--{\bf k}_T^2$, and we do not distinguish between upper
and lower light-cone indices $k^\pm=k_{\pm}$.

The partonic cross-section $d\hat{\sigma}_{\rm
  PRA}$ is:
  \begin{equation}
  d\hat{\sigma}_{\rm PRA}= \frac{\overline{|{\cal A}_{PRA}|^2}}{2Sx_1x_2}\cdot (2\pi)^4 \delta^{(4)}\left(
   q_{1}+ q_{2} - k_1-k_2 \right)  d\Phi(k_1,k_2),\label{eq_partonic}
  \end{equation}
where $d\Phi(k_1,k_2)$ is the element of Lorentz-invariant phase
space for final-state leptons, $2 x_1 x_2 S$ is the appropriate
flux-factor for initial state off-shell partons (see discussion in
Ref.~\cite{PRA:BBbar}).

 The LO unintegrated PDF (unPDF) ${\Phi_{q,\bar q}}(x_{1,2},t_{1,2},\mu^2)$ in Eq.~\ref{eq_factor} is
related with ordinary PDFs of CPM as follows:
\begin{equation}
\Phi_q(x,t,\mu^2)= \frac{T_q(t,\mu^2)}{t}\times
\frac{\alpha_s(t)}{2\pi} \int\limits_x^{1-\Delta} dz
 \frac{x}{z} \left[ P_{qq}(z)
f_q\left(\frac{x}{z},\mu^2 \right)
 +  P_{qg}(z) f_g\left(\frac{x}{z},\mu^2 \right)\right],\label{eq_PDF}
\end{equation}
where $f_{q,g}(x,\mu^2)$ are relevant collinear PDFs, and the
Kimber-Martin-Ryskin cut condition~\cite{KMR, MRW},
$\Delta=\frac{\sqrt{t}}{\sqrt{\mu^2}+\sqrt{t}}$, follows from the
rapidity ordering between the last emission and the hard subprocess.
In Eq.~(\ref{eq_PDF}), $T_q(t,\mu^2)$ is well known Sudakov factor with
boundary conditions $T_q(\mu^2,\mu^2)=T_q(0,\mu^2)=1$, which lead
to the following normalization for unintegrated PDF: $\int\limits_0^{\mu^2}
dt\ \Phi_q(x,t,\mu^2)= xf_q(x,\mu^2)$.

In the PRA, the squared amplitude of the subprocess ($Q\bar Q\to
l^+l^-$)  can be presented as convolution of standard lepton tensor
$L^{\mu\nu}=2[-Q^2 g^{\mu\nu}+2(k_1^\mu k_2^\nu + k_1^\nu k_2^\mu)]$ and the partonic tensor $w_{\mu\nu}^{\rm PRA}$:
\begin{equation}
\overline{|{\cal A}({ Q}\bar { Q}\to
l^+l^-)|^2}=\frac{16\pi^2}{N_c Q^4}\alpha^2e_q^2
L^{\mu\nu}w_{\mu\nu}^{\rm PRA},\nonumber
\end{equation}
where $N_c=3$ and partonic tensor reads
\begin{equation}
w_{\mu\nu}^{\rm PRA}=\frac{1}{4}{\rm tr}\left[\left(\frac{q_2^-}{2}\hat{n}^+\right) \Gamma_\mu(q_1,q_2)\left(\frac{q_1^+}{2}\hat{n}^-\right)
\Gamma_\nu(q_1,q_2) \right], \label{w_mn}
\end{equation}
where factor $1/4$ stands for the averaging over spins of the quark
and antiquark, $\hat{k}=k_\mu \gamma^\mu$ and $\Gamma_\mu(q_1,q_2)$
is the Fadin-Sherman $Q\bar{Q}\gamma$
vertex~\cite{FadinSherman1,FadinSherman2,LipatovVyazovsky}:
\begin{equation}
\Gamma_\mu(q_1,q_2)=\gamma_\mu-\hat{q}_1\frac{n^-_\mu}{q_2^-}-\hat{q}_2\frac{n^+_\mu}{q_1^+}.\label{FS-vertex}
\end{equation}
The QED Ward identity $(q_1+q_2)^\mu \Gamma_\mu(q_1,q_2)=0$ is satisfied by this vertex for any $q_1$ and $q_2$.

\begin{figure}[h]
\centering
\includegraphics[width=0.6\textwidth,clip]{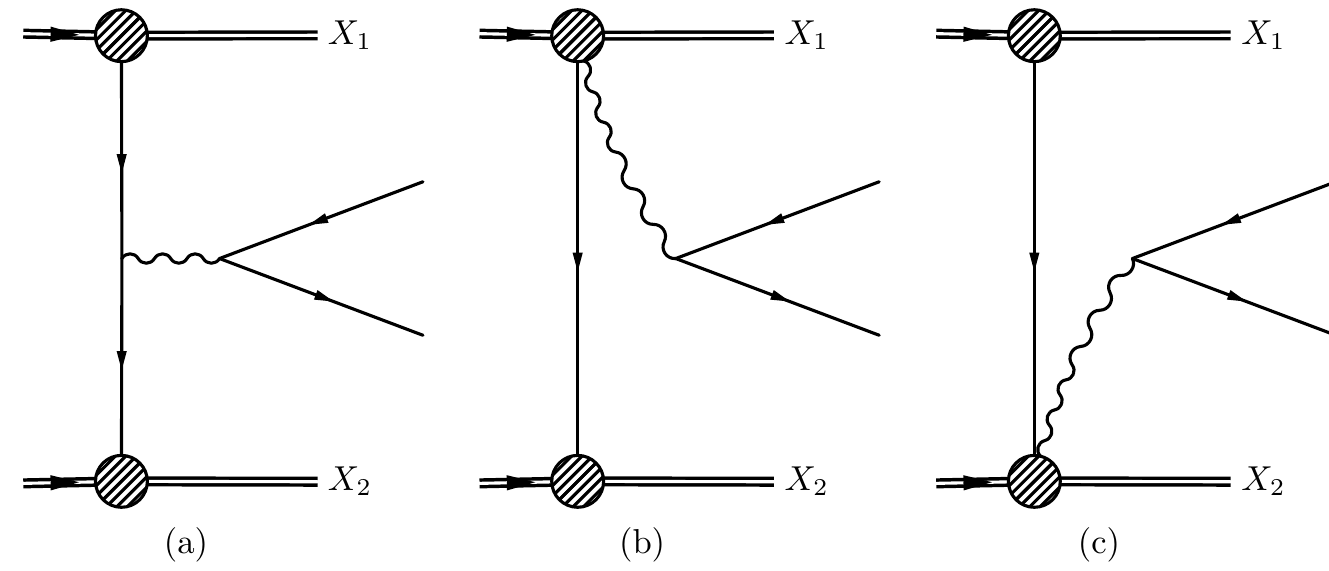}
\caption{Feynman diagrams for the $t$-channel quark-antiquark annihilation subprocess (a), which leads to
the usual parton-model picture, and direct interaction subprocesses (b,c) which are necessary to restore QED gauge invariance of diagram (a).}
\label{fig_diags}       
\end{figure}

 The first term in Eq. (\ref{FS-vertex}) corresponds to the usual $t$-channel quark-antiquark annihilation diagram (a)
in the Fig.~\ref{fig_diags}. While other two(``eikonal'') terms in
Eq. (\ref{FS-vertex}), contain factors $1/q_1^+$ and $1/q_2^-$.
These factors can be understood as a remnants of $s$-channel
propagators in the diagrams where photon interacts with particles
highly separated in rapidity from the lepton pair. More rigorously,
the common lore in high-energy QCD (see e.g. Ref.~\cite{MRK-rev,
Balitsky:WL} and references therein) is, that particles in the
central rapidity region interact with other particles, highly
separated from them in rapidity, as with Wilson lines stretched
along the light-cone. The ``eikonal'' terms in Eq.~(\ref{FS-vertex})
correspond to the coupling of photon with these Wilson lines.
Corrections to this approximation are suppressed by powers of
$e^{-\Delta y}$, where $\Delta y$ is the rapidity gap. In other
words, inclusion of the second and the third terms in
Eq.~(\ref{FS-vertex}) is the simplest possible way to effectively
take into account the diagrams (b) and (c) in the
Fig.~\ref{fig_diags}, where photon interacts directly with the
proton and it's remnants. This approximation assumes only that the
systems $X_1$ and $X_2$ are highly separated in rapidity from the
central region. Rapidity gap between collinear subgraphs exists for $q_T\ll Q$ at the level of {\it leading region} for the Drell-Yan process. This rapidity gap is filled by soft particles emitted from the Glauber gluon exchanges between collinear subgraphs, which does not lead to violation of factorization (See e.g sec. 14.2 and 14.3 in~\cite{Collins}).

\section{Structure functions for DY process in PRA}
\label{spectra} In the notation of Ref.~\cite{Polar-DY-SFs} differential cross section of DY pair production in
collision of non-polarized protons can be written as the combination of helicity
structure functions (SFs):
\begin{eqnarray}
 \frac{d\sigma}{dx_A dx_B d^2{\bf q}_T d\Omega} &=& \frac{\alpha^2}{4Q^2} \Bigl[ F_{UU}^{(1)}\cdot
 \left( 1+\cos^2 \theta \right) + F_{UU}^{(2)}\cdot\left( 1-\cos^2 \theta \right)+\nonumber\\
  &+&F_{UU}^{(\cos \phi)} \cdot \sin(2\theta)\cos\phi +F_{UU}^{(\cos 2\phi)} \cdot \sin^2\theta\cos(2\phi) \Bigr],\nonumber
\end{eqnarray}
were $x_{A,B}=Q e^{\pm Y}/\sqrt{S}$, angles $\theta$ and $\phi$ are defined in the Collins-Soper frame~\cite{CS-frame} and $F_{UU}^{1,2,\cos 2\phi}$
are the helicity SFs at some fixed values of
 $S,q_T=|{\bf q}_{T1}+{\bf q}_{T2}|,x_A, x_B$. With the help of factorization formula (\ref{eq_factor}) SFs can be represented as:
\begin{equation}
F_{UU}^{(1,\ldots)}=\frac{S}{6\pi^2Q_T^4}\int
dt_1\int d\phi_1 \sum_q \Phi_q^p(x_1,t_1,\mu^2)\Phi_{\bar
q}^p(x_2,t_2,\mu^2)\cdot e_q^2 f^{(1,\ldots)},
\end{equation}
where $t_2=({\bf q}_T-{\bf q}_{T1})^2$, $Q_T^2=Q^2+q_T^2$ and $e_q$
is the quark electric charge in units of electron charge.Projecting
the partonic tensor~(\ref{w_mn}) on transverse, longitudinal, single
spin-flip and double spin-flip helicity states of the virtual
photon, one obtains the following expressions for partonic SFs in
PRA~\cite{DY-PRA}:
\begin{eqnarray}
  f_{\rm PRA}^{(1)}=Q^2+\frac{q_T^2}{2},\ f_{\rm PRA}^{(2)}=({\bf q}_{T1}-{\bf q}_{T2})^2, \nonumber \\
f_{\rm PRA}^{(\cos \phi)}=\sqrt{\frac{Q^2}{q_T^2}}({\bf q}_{T1}^2-{\bf q}_{T2}^2),\ f_{\rm PRA}^{(\cos 2\phi)}=\frac{q_T^2}{2}. \label{f_PRA}
\end{eqnarray}

  In the case of collisions of identical target and projectile (e.g. in $pp$-collisions), the SF $F_{UU}^{\cos\phi}$ is
equal to zero in PRA, due to the factor $(t_1-t_2)$ in
Eq.~\ref{f_PRA}. However for collisions of different particles we
expect nonzero value of $F_{UU}^{\cos\phi}$ due to the difference of
transverse-momentum distributions of quarks and antiquarks in the
projectile and in the target\footnote{In the Ref.~\cite{DY-PRA}
partonic coefficient $w_\Delta$ corresponding to the $\cos\phi$
harmonic has been erroneously put to zero. This has no effect on the
plots published in~\cite{DY-PRA}, however now we predict small but
nonzero value of angular coefficient $\mu$, which is still
compatible with NuSea~\cite{NuSea} experimental data for $pD$
collisions within uncertainties. The erratum is in preparation and
will be submitted to the Phys. Rev. D.}.

\section{Quasi-on-shell schemes}
\label{QOS} From the point of view of standard TMD
factorization~\cite{Collins} terms in Eq.~(\ref{FS-vertex}) which
restore the Ward identity for $t_{1,2}\neq 0$ can be viewed as
corrections sub-leading in powers of $q_T/Q$. Therefore it is not
obvious that these terms have significant numerical effect on the
SFs at moderate $q_T<Q$ and especially for $q_T\ll Q$. It is
tempting to say, that the scheme of restoration of gauge-invariance
of partonic tensor is not unique, and all of them should lead to the
same results for SFs at $q_T\ll Q$.

  The simplest way to restore gauge-invariance, retaining the transverse momentum of initial-state partons,
is to artificially put their virtuality to zero on the level of
hard-scattering coefficient. Such hard-scattering coefficient is
just an amplitude of scattering of on-shell partons, which satisfies
the Ward identity automatically. We call such an approach -- {\it
quasi-on-shell (QOS) scheme}.

  Below we will compare the results of PRA with two versions of QOS-scheme. In the Ref.~\cite{Collins} (Sec. 14.5.2)
the hard-scattering coefficient does not depend explicitly on ${\bf
q}_{T1}$ and ${\bf q}_{T2}$, and four-momenta of initial-state
partons, which has been used for the calculation of the partonic
tensor, has been chosen as follows:
\begin{eqnarray}
 ({\tilde{q}_1^{\rm (QOS-1)}})^\mu=\frac{1}{4\kappa}\left(q^+ (\kappa+1) n^\mu_- +  q^- (\kappa-1) n^\mu_+\right) + \frac{q_{T}^\mu}{2}, \nonumber \\
 ({\tilde{q}_2^{\rm (QOS-1)}})^\mu=\frac{1}{4\kappa}\left(q^+ (\kappa-1) n^\mu_- +  q^- (\kappa+1) n^\mu_+\right) + \frac{q_{T}^\mu}{2}, \label{QOS:Collins}
\end{eqnarray}
where $\kappa=\sqrt{Q_T^2/Q^2}$ and $q^\pm=Q_T e^{\pm Y}$, so that
$\tilde{q}_1+\tilde{q}_2=q$ while $\tilde{q}_{1,2}^2=0$. In the
QOS-approximation, the partonic tensor reads:
\[
w^{\rm QOS}_{\mu\nu}=\frac{1}{4}{\rm tr}\left[\hat{\tilde{q}}_2 \gamma_\mu \hat{\tilde{q}}_1\gamma_\nu \right],
\]
and the only nonzero partonic SF, corresponding to the choice (\ref{QOS:Collins}), is $f^{(1)}_{\rm QOS-1}=Q^2$ while $f^{(2)}_{\rm QOS-1}=f^{(\cos\phi)}_{\rm QOS-1}=f^{(\cos 2\phi)}_{\rm QOS-1}=0$ like in CPM.

  To do better, one can try to re-introduce the ${\bf q}_{T1,2}$-dependence into the QOS-scheme. To this end, one adds
the ``small'' light-cone components $q_1^-$ and $q_2^+$ to put
vectors $\tilde{q}_{1,2}$ on-shell:
\begin{eqnarray}
 ({\tilde{q}_1^{\rm (QOS-2)}})^\mu=\frac{1}{2}\left(q_1^+ n^\mu_- +  \frac{{\bf q}_{T1}^2}{q_1^+} n^\mu_+\right) + q_{T1}^\mu, \nonumber \\
 ({\tilde{q}_2^{\rm (QOS-2)}})^\mu=\frac{1}{2}\left(\frac{{\bf q}_{T2}^2}{q_2^-} n^\mu_- +  q_2^- n^\mu_+\right) + q_{T2}^\mu, \label{QOS:2}
\end{eqnarray}
where ``large'' light-cone components are determined from the condition $\tilde{q}_1+\tilde{q}_2=q$ to be
$q_1^+=(Q_T^2+t_1-t_2+\sqrt{D})/(2q^-)$ and $q_2^-=(Q_T^2-t_1+t_2 + \sqrt{D} )/(2q^+)$
 where $D=(Q_T^2-t_1-t_2)^2-4t_1t_2$. Partonic SFs in the new QOS scheme are equal to:
\begin{eqnarray}
&\hspace{-4mm}f^{(1)}_{\rm QOS-2} = Q^2-\frac{({\bf q}_{T1}-{\bf q}_{T2})^2}{2} + \frac{({\bf q}_{T1}^2-{\bf q}_{T2}^2)^2}{2Q_T^2},\ f^{(2)}_{\rm QOS-2} = ({\bf q}_{T1}-{\bf q}_{T2})^2-\frac{({\bf q}_{T1}^2-{\bf q}_{T2}^2)^2}{Q_T^2}, \nonumber \\
&\hspace{-4mm}f^{(\cos\phi)}_{\rm QOS-2}= \sqrt{\frac{Q^2 D}{q_T^2}} \frac{{\bf q}_{T1}^2-{\bf q}_{T2}^2}{Q_T^2},\nonumber \\
&\hspace{-4mm}f^{(\cos 2\phi )}_{\rm QOS-2}=-\frac{({\bf q}_{T1}-{\bf q}_{T2})^2}{2}+\frac{Q^2+Q_T^2}{2Q_T^2}\frac{({\bf q}_{T1}^2-{\bf q}_{T2}^2)^2}{q_T^2}. \label{f-QOS}
\end{eqnarray}

  In this version of QOS-scheme, the coefficients $f_{\rm QOS-2}^{(1)}$, $f_{\rm QOS-2}^{(2)}$ and $f_{\rm QOS-2}^{(\cos\phi)}$
are equal to PRA results at leading power in $|{\bf q}_{T1,2}|/Q$,
however the coefficient $f_{\rm QOS-2}^{(\cos 2\phi)}$ is completely
different from the PRA result. At small ${\bf q}_T={\bf q}_{T1}+{\bf
q}_{T2}$ the first term dominates and this coefficient is negative.

\section{Numerical results and discussion}
\label{num} 
\begin{figure}[h]
\centering
\includegraphics[width=0.5\textwidth,clip]{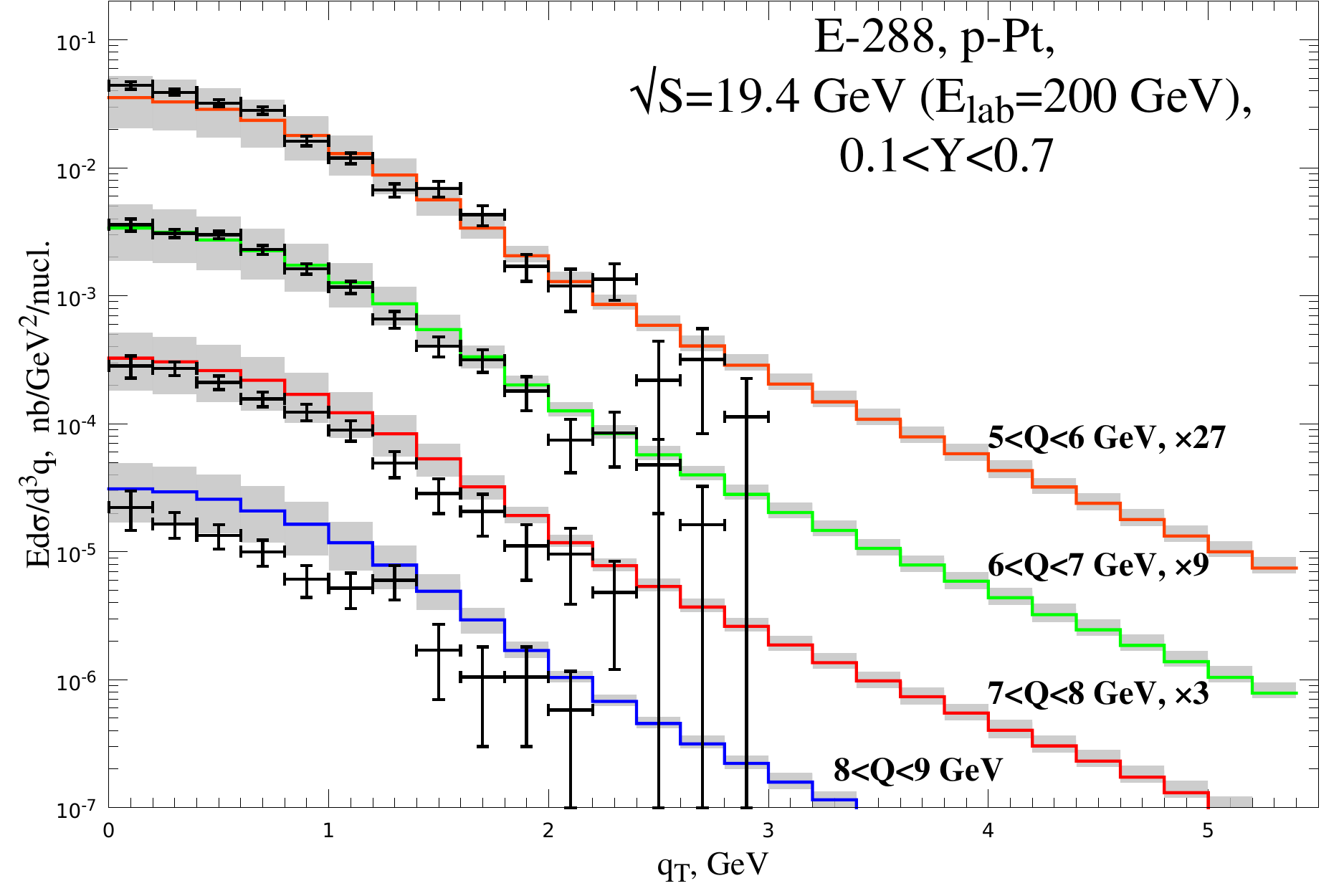}\includegraphics[width=0.5\textwidth,clip]{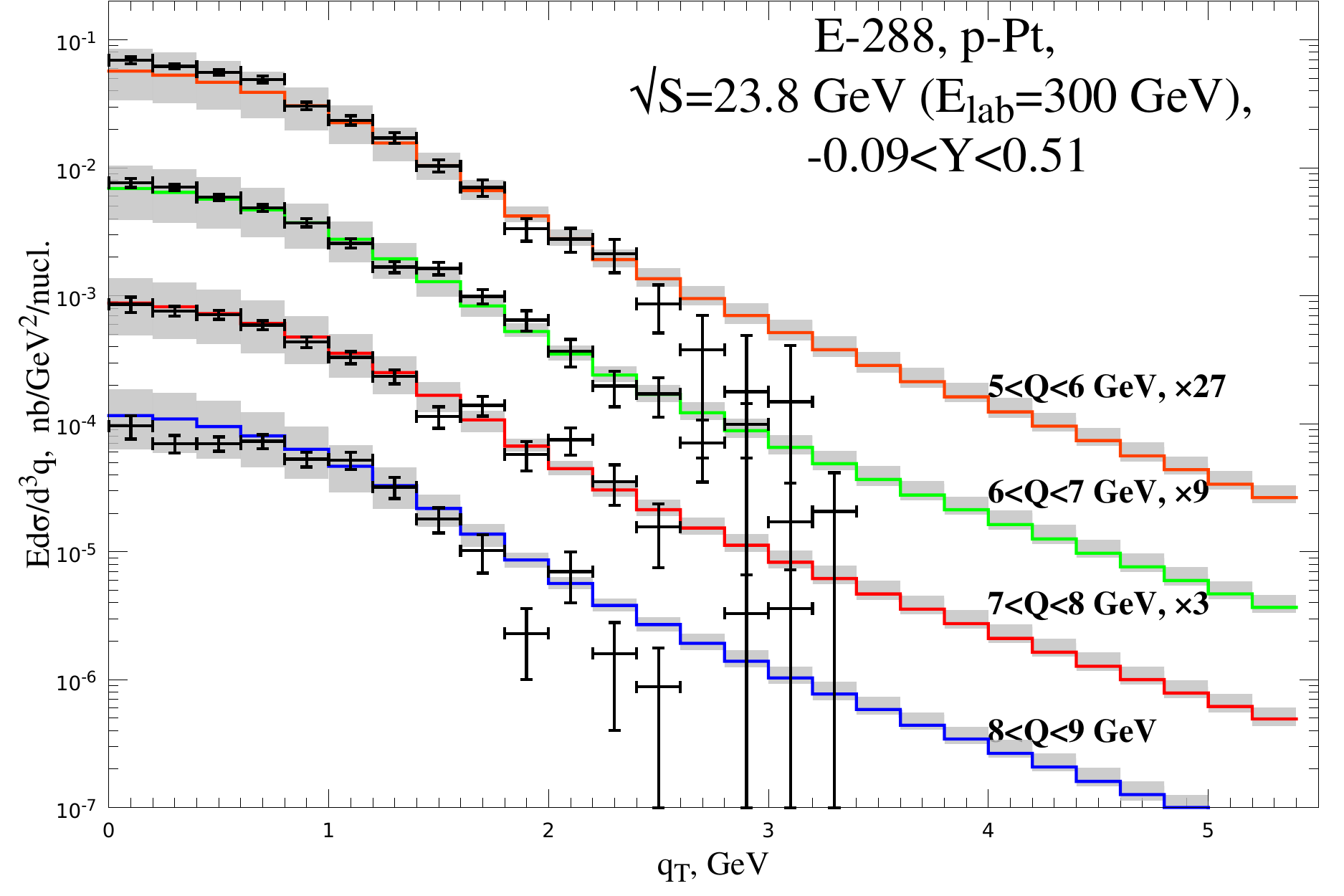}
\caption{Transverse momentum spectra of DY pairs. The histogram
corresponds to calculation in PRA. The data are from the E288
Collaboration~\cite{E-288}, left panel: $\sqrt{S}=19.4$ GeV, $0.1<Y<0.7$;
right panel: $\sqrt{S}=23.8$ GeV, $-0.09<Y<0.51$.}
\label{fig-1}       
\end{figure}

To justify the use of PRA at relatively low $\sqrt{S}=
24$ GeV, which is expected to be achieved during the operation of
NICA collider in the $pp$-collider mode (see
e.g.~\cite{NICA-report}), we compare our numerical results for the
differential cross-section $E d\sigma/d^3{\bf q}$ as a function of
$q_T$ and $Q$ with experimental data of E-288
Collaboration~\cite{E-288}, obtained in the collisions of the proton
beam with platinum fixed target at $\sqrt{S}=19.4$ and $23.8$ GeV
(Fig.~\ref{fig-1}). The KMR unPDF is generated from the LO PDFs
MSTW-2008~\cite{MSTW-2008}.  We use the factorization scale-choice
$\mu_F=\xi Q_T$ and vary $\xi$ in the range $1/2\leq\xi\leq 2$ to
obtain the scale-uncertainty band. The ``$\pi^2$-resummation''
K-factor (see Eq.~(53) in Ref.~\cite{DY-PRA}) is applied to the
cross-section. From the Fig.~\ref{fig-1} one can see, that LO PRA
calculation describes the E-288 data at all values of $Q$ and $Y$
reasonably well.

  Comparison of LO PRA predictions for the $q_T$-dependence of polarization parameters $\lambda$ and $\nu$ with experimental data
of NuSea Collaboration~\cite{NuSea} obtained in the $pp$-collisions
with $\sqrt{S}=39$ GeV is presented in the Ref.~\cite{DY-PRA} and
also demonstrates a good agreement with data. This agreement
justifies our attempt to provide the predictions for helicity SFs
below.

\begin{figure}[h]
\centering
\includegraphics[width=0.5\textwidth,clip]{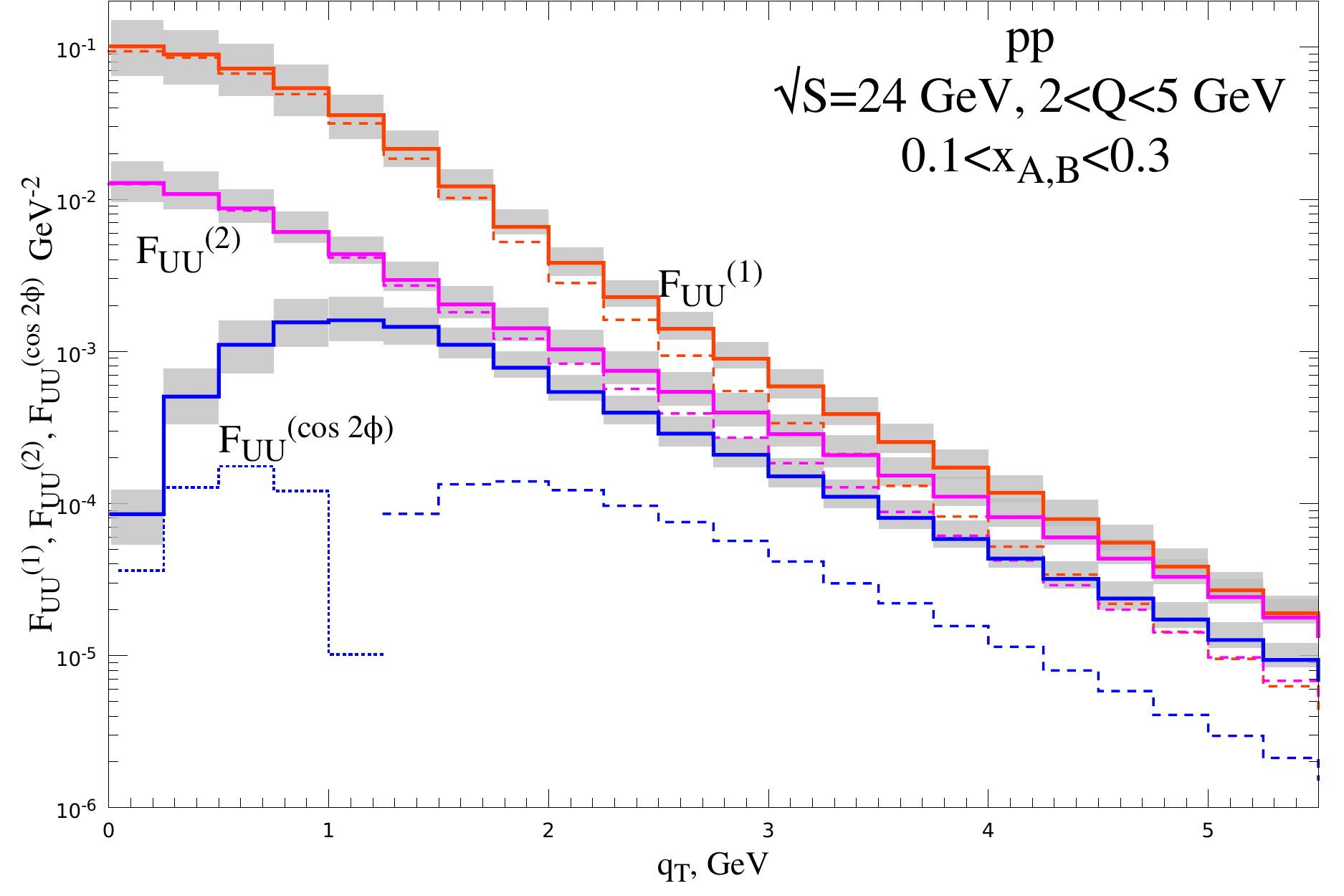}\includegraphics[width=0.5\textwidth,clip]{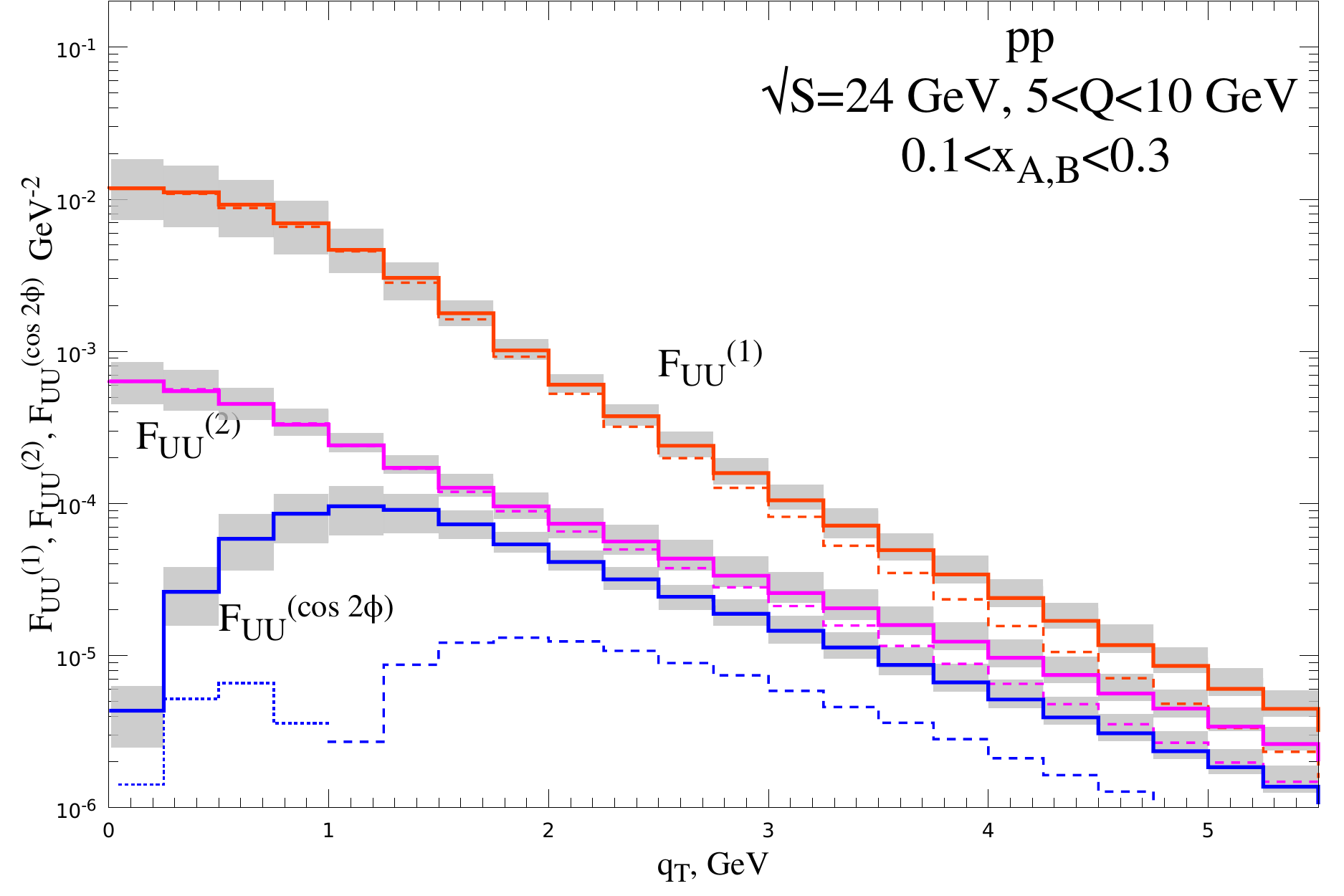}
\caption{Predictions for unpolarized Drell-Yan SFs $F_{UU}^{(1)}$, $F_{UU}^{(2)}$ and $F_{UU}^{(\cos 2\phi)}$ in $pp$-collisions at $\sqrt{S}=24$ GeV.
Solid lines with uncertainty bands -- PRA predictions. Dashed lines -- predictions in the QOS-scheme for the default scale-choice.
Short-dashed line -- plot of the $(-F_{UU}^{(\cos 2\phi)})$ in the QOS scheme, since this SF in QOS scheme is negative at low $q_T$.}
\label{fig-2}       
\end{figure}

 In the Fig.~\ref{fig-2} the PRA predictions for helicity SFs $F_{UU}^{(1,2,\cos 2\phi)}$ are plotted for the case of
$pp$-collisions with $\sqrt{S}=24$ GeV for two bins in the invariant
mass of the pair: $2\leq Q\leq 5$ GeV and $5\leq Q\leq 10$ GeV.
Also, the central lines of predictions of the QOS-scheme, obtained
with the same KMR unPDFs but using the partonic SFs (\ref{f-QOS})
are plotted in the Fig.~\ref{fig-2} together with PRA predictions.

  As expected from the comparison of partonic SFs in the Sec.~\ref{QOS}, the PRA and QOS predictions for $F_{UU}^{(1)}$
and $F_{UU}^{(2)}$ agree for $q_T\ll Q$, however the SFs
$F_{UU}^{(\cos 2\phi)}$ in two approaches differ by more than factor
3 for $q_T>2$ GeV and have different signs for $q_T\to 0$.

  In ``parton-model style'' TMD-factorization~\cite{Polar-DY-SFs}, based solely on the $q\bar{q}$-annihilation picture (diagram (a)
in the Fig.~\ref{fig_diags}), the TMD quark correlators for the case
of unpolarized protons are parametrized in terms of un-polarized
quark distribution $f_1^q(x,{\bf q}_{T}^2)$ and
Boer-Mulders~\cite{B-M_function} function $h_1^{\perp q}(x,{\bf
q}_T^2)$. While former is responsible for the $(1+\cos^2\theta)$
angular dependence and contributes mostly to $F^{(1)}_{UU}$
function, the latter leads to nonzero $F^{(\cos 2\phi)}_{UU}$. The
$(1-\cos^2\theta)$ angular dependence does not arise in
TMD-factorization at leading power in
$q_T^2/Q^2$~\cite{Polar-DY-SFs}. In agreement with this, in PRA the
SF $F_{UU}^{(2)}$ is suppressed by factor $Q^2$ w.r.t.
$F_{UU}^{(1)}$.

 As we have shown above, numerical value of $F_{UU}^{\cos 2\phi}$ even at $q_T^2\ll Q^2$ strongly depends on the details of
the procedure of restoration of gauge-invariance of the
hard-scattering coefficient. On the other hand, in the
TMD-factorization, based on the diagram (a) in the
Fig.~\ref{fig_diags}, the hadronic tensor (e.g. Eq. (73)
in~\cite{Polar-DY-SFs}) does not satisfy Ward identity for $q_T\neq
0$. This raises serious doubts about the Boer-Mulders function as
well-defined physical quantity in this approach.

\section*{Acknowledgements} Authors thank the Ministry of Education and Science of the Russian
Federation for financial support in the framework of the Samara
University Competitiveness Improvement Program among the world's
leading research and educational centers for 2013-2020, the task
number 3.5093.2017/8.9. M.N. acknowledges the support by the
Research Fellowship for postdoctoral researchers of the Alexander
von Humboldt Foundation.

\bibliography{mybibfile}

\begin{thebibliography}{10}
\expandafter\ifx\csname url\endcsname\relax
  \def\url#1{\texttt{#1}}\fi
\expandafter\ifx\csname urlprefix\endcsname\relax\def\urlprefix{URL }\fi
\expandafter\ifx\csname href\endcsname\relax
  \def\href#1#2{#2} \def\path#1{#1}\fi

\bibitem{CSS}
J.~C. Collins, D.~E. Soper, G.~F. Sterman, {Transverse Momentum Distribution in
  Drell-Yan Pair and W and Z Boson Production}, Nucl. Phys. B250 (1985)
  199--224.
\newblock \href {http://dx.doi.org/10.1016/0550-3213(85)90479-1}
  {\path{doi:10.1016/0550-3213(85)90479-1}}.

\bibitem{Collins}
J.~C. Collins, Foundations of perturbative QCD, Cambridge University Press,
  Cambridge, New York, Melbourne, Madrid, Cape Town, Singapore, San Paulo,
  Delhi, Mexico City, 2011.

\bibitem{PRA:BBbar}
A.~V. Karpishkov, M.~A. Nefedov, V.~A. Saleev, {$B{\bar B}$ angular
  correlations at the LHC in parton Reggeization approach merged with
  higher-order matrix elements}, Phys. Rev. D96~(9) (2017) 096019.
\newblock \href {http://arxiv.org/abs/1707.04068} {\path{arXiv:1707.04068}},
  \href {http://dx.doi.org/10.1103/PhysRevD.96.096019}
  {\path{doi:10.1103/PhysRevD.96.096019}}.

\bibitem{MRK-rev}
L.~N. Lipatov, Small x physics in perturbative {QCD}, Phys. Rept. 286 (1997)
  131--198.
\newblock \href {http://dx.doi.org/10.1016/S0370-1573(96)00045-2}
  {\path{doi:10.1016/S0370-1573(96)00045-2}}.

\bibitem{KMR}
M.~A. Kimber, A.~D. Martin, M.~G. Ryskin, Unintegrated parton distributions,
  Phys. Rev. D63 (2001) 114027.
\newblock \href {http://dx.doi.org/10.1103/PhysRevD.63.114027}
  {\path{doi:10.1103/PhysRevD.63.114027}}.

\bibitem{MRW}
G.~Watt, A.~D. Martin, M.~G. Ryskin, Unintegrated parton distributions and
  inclusive jet production at hera, Eur. Phys. J. C31 (2003) 73--89.
\newblock \href {http://dx.doi.org/10.1140/epjc/s2003-01320-4}
  {\path{doi:10.1140/epjc/s2003-01320-4}}.

\bibitem{FadinSherman1}
V.~S. Fadin, V.~E. Sherman, Fermion {Reggeization} in non-abelian gauge
  theories, JETP Lett. 23 (1976) 599.

\bibitem{FadinSherman2}
V.~S. Fadin, V.~E. Sherman, Fermion exchange processes processes in non-abelian
  gauge theories, Sov. Phys. JETP 45 (1977) 861.

\bibitem{LipatovVyazovsky}
L.~Lipatov, M.~Vyazovsky, Quasi-multi-regge processes with a quark exchange in
  the t-channel, Nucl. Phys. B596 (2001) 399.
\newblock \href {http://arxiv.org/abs/0009340} {\path{arXiv:0009340}}, \href
  {http://dx.doi.org/10.1016/S0550-3213(00)00709-4}
  {\path{doi:10.1016/S0550-3213(00)00709-4}}.

\bibitem{Balitsky:WL}
I.~Balitsky, G.~A. Chirilli, {Rapidity evolution of Wilson lines at the
  next-to-leading order}, Phys. Rev. D88 (2013) 111501.
\newblock \href {http://arxiv.org/abs/1309.7644} {\path{arXiv:1309.7644}},
  \href {http://dx.doi.org/10.1103/PhysRevD.88.111501}
  {\path{doi:10.1103/PhysRevD.88.111501}}.

\bibitem{Polar-DY-SFs}
S.~Arnold, A.~Metz, M.~Schlegel, {Dilepton production from polarized hadron
  hadron collisions}, Phys. Rev. D79 (2009) 034005.
\newblock \href {http://arxiv.org/abs/0809.2262} {\path{arXiv:0809.2262}},
  \href {http://dx.doi.org/10.1103/PhysRevD.79.034005}
  {\path{doi:10.1103/PhysRevD.79.034005}}.

\bibitem{CS-frame}
J.~C. Collins, D.~E. Soper, {Angular Distribution of Dileptons in High-Energy
  Hadron Collisions}, Phys. Rev. D16 (1977) 2219.
\newblock \href {http://dx.doi.org/10.1103/PhysRevD.16.2219}
  {\path{doi:10.1103/PhysRevD.16.2219}}.

\bibitem{DY-PRA}
M.~A. Nefedov, N.~N. Nikolaev, V.~A. Saleev, {Drell-Yan lepton pair production
  at high energies in the Parton Reggeization Approach}, Phys. Rev. D87~(1)
  (2013) 014022.
\newblock \href {http://arxiv.org/abs/1211.5539} {\path{arXiv:1211.5539}},
  \href {http://dx.doi.org/10.1103/PhysRevD.87.014022}
  {\path{doi:10.1103/PhysRevD.87.014022}}.

\bibitem{NuSea}
L.~Y. Zhu, et~al., {Measurement of Angular Distributions of Drell-Yan Dimuons
  in p + p Interactions at 800-GeV/c}, Phys. Rev. Lett. 102 (2009) 182001.
\newblock \href {http://arxiv.org/abs/0811.4589} {\path{arXiv:0811.4589}},
  \href {http://dx.doi.org/10.1103/PhysRevLett.102.182001}
  {\path{doi:10.1103/PhysRevLett.102.182001}}.

\bibitem{NICA-report}
I.~A. Savin, et~al., {Spin Physics Experiments at NICA-SPD with polarized
  proton and deuteron beams}, EPJ Web Conf. 85 (2015) 02039.
\newblock \href {http://arxiv.org/abs/1408.3959} {\path{arXiv:1408.3959}},
  \href {http://dx.doi.org/10.1051/epjconf/20158502039}
  {\path{doi:10.1051/epjconf/20158502039}}.

\bibitem{E-288}
A.~S. Ito, et~al., {Measurement of the Continuum of Dimuons Produced in
  High-Energy Proton - Nucleus Collisions}, Phys. Rev. D23 (1981) 604--633.
\newblock \href {http://dx.doi.org/10.1103/PhysRevD.23.604}
  {\path{doi:10.1103/PhysRevD.23.604}}.

\bibitem{MSTW-2008}
A.~D. Martin, W.~J. Stirling, R.~S. Thorne, G.~Watt, {Parton distributions for
  the LHC}, Eur. Phys. J. C63 (2009) 189--285.
\newblock \href {http://arxiv.org/abs/0901.0002} {\path{arXiv:0901.0002}},
  \href {http://dx.doi.org/10.1140/epjc/s10052-009-1072-5}
  {\path{doi:10.1140/epjc/s10052-009-1072-5}}.

\bibitem{B-M_function}
D.~Boer, P.~J. Mulders, {Time reversal odd distribution functions in
  leptoproduction}, Phys. Rev. D57 (1998) 5780--5786.
\newblock \href {http://arxiv.org/abs/hep-ph/9711485}
  {\path{arXiv:hep-ph/9711485}}, \href
  {http://dx.doi.org/10.1103/PhysRevD.57.5780}
  {\path{doi:10.1103/PhysRevD.57.5780}}.

\end{thebibliography}

\end{document}